# An Impossibility Result for Truthful Combinatorial Auctions with Submodular Valuations

Shahar Dobzinski
Department of Computer Science
Cornell Unversity
`shahar@cs.cornell.edu`

September 27, 2018

**Abstract**

We show that every universally truthful randomized mechanism for combinatorial auctions with submodular valuations that provides $m^{\frac{1}{2}-\epsilon}$ approximation to the social welfare and uses value queries only must use exponentially many value queries, where $m$ is the number of items. In contrast, ignoring incentives there exist constant ratio approximation algorithms for this problem. Our approach is based on a novel *direct hardness* approach and completely skips the notoriously hard characterization step. The characterization step was the main obstacle for proving impossibility results in algorithmic mechanism design so far.

We demonstrate two additional applications of our new technique: (1) an impossibility result for universally-truthful polynomial time flexible combinatorial public projects and (2) an impossibility result for truthful-in-expectation mechanisms for exact combinatorial public projects. The latter is the first result that bounds the power of polynomial-time truthful in expectation mechanisms in any setting.

# 1 Introduction

This paper attempts to answer one of the earliest open questions in Algorithmic Mechanism Design: is there a truthful computationally-efficient mechanism for combinatorial auctions with submodular bidders that provides a constant approximation ratio?

In a combinatorial auction there is a set $M$, $|M| = m$ of items, and a set $N = \{1, 2, \ldots, n\}$ of bidders. Each bidder $i$ has a valuation function $v_i : 2^M \to \mathbb{R}^+$, which is normalized ($v_i(\emptyset) = 0$) and non-decreasing. An important special case is when each valuation is *submodular*: for every item $j$ and bundles $S$ and $T$, $S \subseteq T$, $v(S \cup \{j\}) - v(S) \geq v(T \cup \{j\}) - v(T)$. The definition captures valuations that exhibit "decreasing marginal utilities". The goal is to maximize the social welfare, i.e., to find an allocation $(S_1, \ldots, S_n)$ that maximizes $\Sigma_i v_i(S_i)$. As in previous work, we would like our algorithms to run in time polynomial in the natural parameters of the problems, $n$ and $m$. Since the valuation function is an object of exponential size, we assume that each valuation $v$ is given to us as a black box that can only answer *value queries*: given $S$, return the value of $v(S)$.

The main interest of this paper is in incentive-compatible algorithms that handle the selfish behavior of the bidders. We are interested in designing *truthful* algorithms in which the profit-maximizing strategy of each bidder is to reveal his true valuation (i.e., truthfully answer the queries).

the problem has received a lot of attention, even from a pure optimization point of view, completely ignoring incentives. The most notable result here is Vondrak's celebrated algorithm [27] that provides an approximation ratio of $\frac{e}{e-1}$, improving over the 2-approximation of the greedy algorithm [21]. This ratio is the best possible with a polynomial number of value queries [18, 23]. While value queries are widely used in the design of algorithms for other optimization scenarios that involve submodular functions (see, e.g., [17, 16, 14]), back in the combinatorial auctions setting, other algorithms guarantee improved approximation ratios using the stronger *demand queries* (given prices $p_1, \ldots, p_m$, return a bundle that maximizes $v(S) - \Sigma_{j \in S} p_j$). The state of the art in this setting is an $(\frac{e}{e-1} - 10^{-4})$-approximation algorithm [15], an improvement over the $\frac{e}{e-1}$-approximation algorithm of [12].

Much less is known regarding the design of truthful algorithms for this problem. The VCG mechanism is a truthful algorithm for the problem, but requires computing the optimal solution and thus is not computable in polynomial time. The best known polynomial time deterministic algorithm provides a poor approximation ratio of $O(\sqrt{m})$ [10]. Whether this ratio is the best possible with deterministic truthful polynomial-time algorithms is the subject of the current paper. If we provide the algorithm designer with more power and allow the use of *both* randomization and the strictly more powerful demand queries, an $O(\log m \log \log m)$-truthful approximation algorithm exists [6, 11]. Unfortunately, despite all progress made over the years, the algorithmic mechanism design community is unable to answer the question posed by Lehmann, Lehmann, and Nisan [21] back in 2001: is there a truthful polynomial-time $O(1)$-approximation algorithm for combinatorial auctions with submodular bidders?

## 1.1 Previous Technique: Characterize and Optimize

Roughly speaking, problems in Algorithmic Mechanism Design are either single parameter or multi-parameter. Single parameter problems, where the private information of each player consists of essentially one number, are quite well understood: an algorithm is truthful if and only if it is monotone (see [24]). This characterization gives rise to many truthful algorithms with approximation ratios that match what is achievable by the best non-truthful polynomial time mechanisms (e.g., [22, 2, 4]).

Combinatorial auctions with submodular bidders belong to the harder class of multi-parameter problems. In this class, the private information of each player consists of more than one parameter (for example, in combinatorial auctions the private information of a bidder consists of exponentially many values of bundles). Since the current best approximation ratios achievable by truthful polynomial time mechanisms are usually quite far from what can be obtained from a pure algorithmic point of view that ignores incentives, great effort was and is invested in proving impossibility results. The main obstacle in proving these impossibilities is the hardness of obtaining useful characterizations for multi-parameter domains. Specifically, all known impossibility results on the power of computationally-efficient truthful mechanisms are proved using the following two-stage paradigm:

1. **Characterize** all truthful mechanisms for the setting, ignoring computational issues.

2. **Optimize** over all truthful algorithms: i.e., show a lower bound on the approximation ratio of the best computationally efficient mechanism characterized in the previous step.

This paradigm was quite successful in obtaining impossibilities for problems with "full dimensionality" [19, 13, 25]: in the first characterization step, it is shown that *all* truthful mechanisms for the problems are VCG-based (a slight generalization of the VCG mechanism), regardless of their approximation ratio – thereby extending Roberts' theorem [26][1]. The second optimization step shows that VCG-based algorithms cannot provide a good approximation ratio in polynomial time.

For combinatorial auctions with submodular valuations, the optimization step was accomplished in [8] where it was shown that every VCG-based $m^{\frac{1}{6}}$-approximation mechanism requires exponential communication. However, completing the characterization step is notoriously hard for auction domains and, in general, domains that do not exhibit externalities: in these domains it is easy to construct truthful mechanisms that are not VCG-based (but these mechanisms can guarantee at best a trivial approximation ratio). Furthermore, as [9] shows, there are non-VCG-based mechanisms that guarantee arbitrarily good approximation ratios[2]! Till now, [9] is the only example of a successful characterization of truthful mechanisms for a multi-parameter auction domain, and even there the extra assumption of scalability is needed. Moreover, the characterization of multi-unit auctions of [9] holds only for two bidders. While this suffices for obtaining an optimal inapproximability result for multi-unit auctions, an optimal result for combinatorial auctions probably requires characterization of mechanisms for many bidders. This task seems to be quite difficult: we do not even have a good conjecture of what the class of mechanisms with good approximation ratios might be[3].

## 1.2 Our Results: Impossibilities via Direct Hardness

This paper introduces a simple technique for bounding the power of truthful mechanisms. The technique is very different from the characterize-and-optimize approach, and in particular does not require obtaining characterizations of truthful mechanisms at all. The starting point is the *taxation principle*: consider some player $i$, and fix the valuations of all other players. According to the taxation principle, in a truthful algorithm each bundle $S$ has a price $p_S$ (possibly $\infty$) and bidder $i$ is assigned the bundle that maximizes his profit $v(S) - p_S$. We call this set of bundles and prices the *menu* of player $i$. We show that in any algorithm that provides a good approximation there exist valuations such that some bidder faces a "large" menu with a "nice" structure. We then prove that selecting the profit maximizing bundle in the menu – a must according to the taxation principle – requires exponentially many value queries. This leads us to the statement of our main result:

**Theorem:** Let $A$ be a randomized universally truthful mechanism for combinatorial auctions with submodular bidders that provides an approximation ratio of $m^{\frac{1}{2}-\epsilon}$, for some constant $\epsilon > 0$. Then, A makes exponentially many value queries.

Notice that our result holds not only for deterministic mechanisms but also for universally truthful mechanisms (i.e., a probability distribution over truthful deterministic mechanisms). This is yet another benefit of skipping the characterization step and using our direct hardness approach.

---
[1] Roberts [26] shows that if the domain of valuations is unrestricted then every truthful algorithm is VCG-based.
[2] But these mechanisms (for multi-unit auctions) are not computationally efficient.
[3] The problem is even more acute for *randomized* mechanisms: characterizations of truthful randomized mechanisms are probably an impossible task using our current techniques.

### 1.2.1 Flexible Combinatorial Public Projects

We then proceed to show the applicability of our techniques in other domains. Papadimitriou et al. [25] presented the combinatorial public project problem. Similarly to a combinatorial auction, there are $m$ items and $n$ players with monotone and normalized submodular valuations. Unlike combinatorial auctions the goal is to find a *single* bundle $S$ of size *exactly $k$* that maximizes $\Sigma_i v_i(S)$. A simple greedy algorithm provides an approximation ratio of $\frac{e}{e-1}$ for this problem ignoring incentives issues.

Papadimitriou et al. use the characterize-and-optimize approach to show a lower bound of $m^{\frac{1}{2}-\epsilon}$ on the approximation ratio of truthful polynomial time algorithms: they first show that all truthful algorithms for the problem are VCG-based, and then that VCG-based algorithms cannot provide a good approximation ratio in polynomial time. However, a natural relaxation of the problem allows outputting bundles of size *at most $k$* (the *flexible* model) rather than bundles of size exactly $k$ (the *exact* model). While this relaxation is useless from a pure algorithmic point of view, since the valuations are monotone, there might be truthful non-VCG based mechanisms in the flexible domain, thus bypassing the characterization and impossibility result of [25][4]. To the very least, characterizing truthful mechanisms in the flexible model seem to require new techniques. Our direct hardness approach allows us to ignore all these complications and obtain the following:

**Theorem:** Let $A$ be a randomized universally truthful mechanism for flexible combinatorial public projects that provides an approximation ratio of $m^{\frac{1}{2}-\epsilon}$, for some constant $\epsilon > 0$. Then, A makes exponentially many value queries.

### 1.2.2 Hardness of Truthful in Expectation Mechanisms

As there has been only limited success in designing powerful deterministic and universally truthful mechanisms for many domains, there is a line of research [1, 20, 5, 7] that advocates the use of a relaxed notion of truthfulness, truthfulness in expectation. In a truthful-in-expectation mechanism, truth telling maximizes the *expected* profit, where the expectation is taken over the internal random coins of the algorithm. Truthfulness in expectation is a reasonable relaxation of deterministic truthfulness, but one should keep in mind that it should be used only if bidders are known to be risk neutral and not, for example, risk averse (in contrast to universally truthful mechanisms, see [11] for a discussion). It is known that truthfulness in expectation is strictly stronger than deterministic truthfulness in some settings [7]. Can truthfulness in expectation be the remedy for all pitfalls of deterministic truthfulness? Unfortunately, we give a negative answer:

**Theorem:** Let $A$ be a randomized truthful-in-expectation mechanism for exact combinatorial public projects that provides an approximation ratio of $m^{\frac{1}{2}-\epsilon}$, for some constant $\epsilon > 0$. Then, A makes exponentially many value queries.

This is the first lower bound on the power of polynomial time truthful-in-expectation mechanisms in any setting. We again prove an impossibility without a characterization[5]. Yet, exact combinatorial public projects have a somewhat artificial flavor in our opinion, especially in a randomized setting[6]. Unfortunately, we currently do not know how to extend our result, and whether there exists an efficient truthful-in-expectation mechanism with a good approximation ratio in the flexible model remains an open question.

---

[4] A similar phenomenon exists in multi-unit auctions, where truthful algorithms that always allocate all items must be VCG-based [13, 19], but without this extra condition the triage mechanisms of [9] are non-VCG based truthful algorithms that provide a good approximation ratio.

[5] Nevertheless, it is extremely interesting to obtain a characterization of truthful-in-expectation mechanisms in any multi-parameter setting. Even in Roberts' setting [26], where the valuations are completely unrestricted, such a characterization is not known!

[6] Randomized mechanisms can sometimes take advantage of not allocating all items. See [7] for an example.

## 1.3 Open Questions

This paper shows that every universally truthful randomized mechanism for combinatorial auctions with submodular valuations with an approximation ratio of $m^{\frac{1}{2}-\epsilon}$ makes an exponential number of value queries. This was achieved by introducing a novel approach that allows proving hardness without characterization. Nevertheless, a full characterization of truthful mechanisms with good approximation ratio remains an important question, even ignoring computational implications.

If demand queries are allowed, there exists a *randomized* universally-truthful $O(\log m \log \log m)$-approximation algorithm [6]. Is there an $m^{\frac{1}{2}-\epsilon}$ *deterministic* algorithm that uses a polynomial number of *demand* queries? A truthful-in-expectation $O(1)$-approximation mechanism that uses demand queries, or even a universally truthful one? Another open question is to prove hardness results that are based on computational complexity rather than on concrete complexity, for, say, the budget additive case (see [25, 3]). These questions remain open, but we do believe that a refinement of our direct-hardness technique might be capable of making significant progress in providing answers.

**Paper Organization**

Section 2 is the preliminaries section. Section 3 contains our main result: an impossibility result for truthful polynomial time combinatorial auctions with submodular valuations. The subject of Section 4 is an impossibility result for combinatorial public projects. Finally, in Section 5 we discuss truthful in expectation mechanisms for exact combinatorial public projects.

## 2 Preliminaries

### 2.1 The Settings

#### 2.1.1 Combinatorial Auctions with Submodular Valuations

In a combinatorial auction there is a set $M$, $|M| = m$ of items, and a set $N = \{1, 2, \ldots, n\}$ of bidders. Each bidder $i$ has a valuation function $v_i : 2^M \to \mathbb{R}^+$, which is normalized ($v_i(\emptyset) = 0$) and non-decreasing. We assume that the valuations are submodular: a valuation $v$ is *submodular* if it exhibits decreasing marginal utilities, $v(S \cup \{j\}) - v(S) \geq v(T \cup \{j\}) - v(T)$, for every item $j$ and bundles $S, T$, $S \subseteq T$. Equivalently, $v(S) + v(T) \geq v(S \cup T) + v(S \cap T)$, for every two bundles $S$ and $T$.

Let $V$ be the set of all submodular valuations. An allocation of the items $\vec{S} = (S_1, \ldots, S_n)$ is a vector of pairwise disjoint of subsets of $M$. Let $\mathbb{S}$ be the set of all allocations. The goal is to find an allocation that maximizes the welfare: $\Sigma_i v_i(S_i)$. The valuations are given as black boxes. We assume that the black box $v$ is accessed only via value queries: given a bundles $S$, what is $v(S)$. We want our algorithms to make a polynomial number (in $n$ and $m$) of value queries to the black boxes.

#### 2.1.2 Combinatorial Public Projects

In a combinatorial public project, as in combinatorial auctions, we also have a set $M$, $|M| = m$ of items, and a set $N = \{1, 2, \ldots, n\}$ of bidders. Similarly, each bidder $i$ has a valuation function $v_i : 2^M \to \mathbb{R}^+$, which is normalized ($v_i(\emptyset) = 0$), non-decreasing and submodular. The valuations are given as black boxes that can only answer value queries. In *exact combinatorial public projects* (this is the model defined in [25]) the goal is to find a bundle $S$ of size *exactly* $k$ that maximizes $\Sigma_i v_i(S)$. In *flexible combinatorial public projects* we are allowed to output $S$ of size *at most* $k$ that maximizes $\Sigma_i v_i(S)$. We are interested in algorithms that make a polynomial number (in $n$ and $m$) of value queries.

## 2.2 Truthfulness

The reader is referred to [24] for the (standard) proofs missing in this subsection. An $n$-bidder mechanism is a pair $(A, p)$ where $A : V^n \to \mathbb{S}$ and $p = (p^{(1)}, \cdots, p^{(n)})$, where for each $i$, $p^{(i)} : V^n \to \mathbb{R}$.

**Definition 2.1** *Let $(A, p)$ be a deterministic mechanism. $(A, p)$ is* truthful *if for all $i$, all $v_i, v_i'$ and all $v_{-i}$ we have that $v_i(A(v_i, v_{-i})_i) - p^{(i)}(v_i, v_{-i}) \geq v_i(A(v_i', v_{-i})_i) - p^{(i)}(v_i', v_{-i})$.*

It is well known that an algorithm (for combinatorial auctions or combinatorial public projects) is truthful if and only if each bidder is presented with a payment for each bundle $T$ that does not depend on bidder $i$'s valuation (i.e., $p^{(i)} : V^{n-1} \to \mathbb{R}$). Denote this payment by $p_T^{(i)}(v_{-i})$. Each bidder is allocated a bundle that maximizes his profit: $v_i(T) - p_T^{(i)}(v_{-i})$ (this is called the "taxation principle" – we will sometimes say that these payments are *induced* by $v_{-i}$).

**Definition 2.2 (Menu)** *Fix some algorithm $A$. The* menu *of $i$ given $v_{-i}$ in $A$ is*

$$\mathcal{R}_{v_{-i}} = \{S | \exists v_i \text{ s.t. } A(v_i, v_{-i})_i = S\}$$

### 2.2.1 Randomized Mechanisms

**Definition 2.3** *$(A, p)$ is* universally truthful *if it is a probability distribution over truthful deterministic mechanisms.*

**Definition 2.4** *$(A, p)$ is* truthful in expectation *if for all $i$, all $v_i, v_i'$ and all $v_{-i}$ we have that $E[v_i(A(v_i, v_{-i})_i) - p(v_i, v_{-i})] \geq E[v_i'(A(v_i', v_{-i})_i) - p_i(v_i', v_{-i})]$, where the expectation is over the internal random coins of the algorithm.*

## 2.3 Chernoff Bounds

We will use the following version of the chernoff bounds multiple times.

**Proposition 2.5 (Chernoff)** *Let $X_1, ... X_m$ be independent random variables that take values in $\{0, 1\}$, such that for all $i$, $\Pr[X_i = 1] = p$ for some $p$. Then, the following holds, for $0 \leq \epsilon \leq 1$:*

1. $\Pr[\Sigma_i X_i > (1 + \epsilon)pm] \leq e^{-pm\epsilon^2}$
2. $\Pr[\Sigma_i X_i < (1 - \epsilon)pm] \leq e^{-pm\epsilon^2}$

# 3 The Main Result: Combinatorial Auctions with Submodular Valuations

We start with proving a lower bound on deterministic mechanisms. In the appendix we discuss how to extend the lower bound to randomized universally truthful algorithms (Theorem A.1).

**Theorem 3.1** *Let $A$ be a deterministic truthful $\frac{n}{10}$-approximation mechanism for combinatorial auctions with submodular valuations. Then, $A$ makes at least $\frac{e^{\frac{m}{n^2}}}{10n^2 \cdot m^6}$ value queries.*

In particular, for any constant $\epsilon > 0$ and $n = m^{\frac{1}{2} - \epsilon}$, we get that $A$ must make exponential number of value queries to achieve an approximation ratio of $O(m^{\frac{1}{2} - \epsilon})$.

The proof shows that for some $v_i$ and some valuations of the other bidders $v_{-i}$, finding the bundle that maximizes the profit of bidder $i$ $v_i(S) - p_{v_{-i}(S)}$ requires exponential number of value queries. The proof is divided into two parts. In the first part (Section 3.1) we show that there are valuations $v_{-i}$ that induce a submenu with "nice" properties. In the second part (Section 3.2) we use the submenu to define a valuation $v_i$ of bidder $i$ such that finding the profit-maximizing bundle for $v_i$ requires exponential number of value queries.

Specifically, the first step shows that for some $v_{-i}$ the menu of bidder $i$ is exponentially large. This by itself is not enough; the profit-maximizing bundle may be found with only a polynomial number of value queries even in exponentially large menus. Therefore, we find a "large" submenu where the bundles' prices are "almost the same" with the additional property that if a bundle $T$ is in the submenu, then every other bundle $U$ in the menu that contains $T$ has a "significantly" higher price. These two properties, together with other easier-to-show properties, enable us to construct a valuation $v_i$ for which finding the profit-maximizing bundle requires exponentially many value queries.

**Definition 3.2 (Structured Submenu)** *A set $\mathcal{S} \subseteq \mathcal{R}_{v_{-i}}$ is* structured *if*

- *For each $S, S' \in \mathcal{S}$: $|p_S(v_{-i}) - p_{S'}(v_{-i})| \leq \frac{1}{m^5}$.*
- *For all $S, T$ such that $S \in \mathcal{S}$, $T \in \mathcal{R}_{v_{-i}}$ and $T$ strictly contains $S$: $p_T(v_{-i}) - p_S(v_{-i}) \geq \frac{1}{m^3}$.*
- *For all $S \in \mathcal{S}$: $p_S(v_{-i}) \leq m$.*
- *For each $S, S' \in \mathcal{S}$: $|S| = |S'|$.*

## 3.1 Existence of Exponentially Large Structured Submenus

**Lemma 3.3** *Let $A$ be a truthful $\frac{n}{10}$-approximation mechanism for combinatorial auctions with submodular valuations. Then, there exists $v_{-i}$, $\mathcal{S}$, $|\mathcal{S}| \geq \frac{e^{\frac{m}{n^2}}}{10n^2 \cdot m^6}$, such that $\mathcal{S} \subseteq \mathcal{R}_{v_{-i}}$ is a structured submenu.*

The proof makes use of the following class of valuations:

**Definition 3.4** *A valuation $v$ is called* polar additive *if both of the following conditions hold:*

- *$v$ is additive.*
- *For each item $j$ either $v(\{j\}) = 1$ or $v(\{j\}) = \frac{1}{m^3}$.*

We show that there exists $v_{-i}$ that consists of polar additive valuations only, and that the induced menu of $v_{-i}$ contains a structured submenu of at least the specified size. We use the probabilistic method to prove the existence of such $v_i$. The valuation $v_i$ of bidder $i$ is constructed as follows: for each item $j$, set independently at random $v_i(j) = 1$ with probability $p = \frac{1}{n}$, or $v_i(j) = \frac{1}{m^3}$ with probability $1 - p$. We call valuations constructed this way *random*. We say that item $j$ is *demanded* by bidder $i$ if $v_i(\{j\}) = 1$.

**Definition 3.5** *Fix bidder $i$ and $v_{-i}$, where each $v \in v_{-i}$ is polar additive. Let $\mathcal{S}_{v_{-i}} = \{S | \exists \text{ a polar-additive valuation } v_i \text{ s.t. } A(v_i, v_{-i})_i = S\}$.*

**Claim 3.6** *Fix $v_{-i}$, and let $S \in \mathcal{S}_{v_{-i}}$. Then,*

1. *$p_{v_{-i}}(S) \leq m$.*
2. *For each $S \subseteq T$ such that $T \in \mathcal{R}_{v_{-i}}$ we have that $p_{v_{-i}}(T) \geq p_{v_{-i}}(S) + \frac{1}{m^3}$.*

**Proof:** The first property holds since otherwise bidder $i$ with valuation $v_i$ has negative profit for $S$ and thus prefers the empty bundle (which has a profit of zero). The second property holds since the marginal value of every item in a polar additive valuation is at least $\frac{1}{m^3}$. Thus, if the price difference between $S$ and $T$ is less than $\frac{1}{m^3}$, then $S \notin \mathcal{R}_{v_{-i}}$. □

**Claim 3.7** $\Pr[\text{the number of items demanded by at least one bidder} > (1 - 1.01(1-p)^n)m] \leq e^{-\frac{m}{300}}$.

**Proof:** Fix some item $j$. The probability that this item is demanded by no bidder is exactly $(1-p)^n$. By the chernoff bounds:

$$\Pr[\text{the number of items demanded by at least one bidder} > (1 - 1.01(1-p)^n)m] \leq e^{-\frac{(1-p)^n m}{100}} < e^{-\frac{m}{300}}$$

□

**Claim 3.8** Fix some bundle $S$, $|S| > \frac{pm}{n}$. Let $v_i$ be a random polar-additive valuation. With probability at least $1 - e^{-p|S|}$, $v_i(S) \leq 2p|S| + \frac{1}{m^2}$.

**Proof:** The probability that item $j \in S$ is demanded by bidder $i$ is $p$. By the chernoff bounds, the probability that more than $2p|S|$ of the items will be demanded by bidder $i$ is at most $e^{-p|S|}$. The contribution of the items that are not demanded by $i$ is at most $m \cdot \frac{1}{m^3} = \frac{1}{m^2}$. Therefore, in this case $v_i(S) \leq 2p|S| + \frac{1}{m^2}$. □

**Claim 3.9** There exist bidder $i$ and $v_{-i}$ such that $|\mathcal{S}_{v_{-i}}| > \frac{e^{\frac{m}{n^2}}}{10n^2}$.

**Proof:** Consider an instance where each $v_i$ is random. Denote by $O$ the event in which the optimal solution has value of at least $m(1 - 1.01(1-p)^n) > m(1 - \frac{1.01}{e})$. By Claim 3.7, $\Pr[O] \geq 1 - e^{-\frac{m}{300}}$.

For each $i$ and bundle $S \in \mathcal{S}_{v_{-i}}$, $|S| > \frac{pm}{n}$, denote by $C_S^i$ the event in which $v_i(S) \leq 2p|S_i| + \frac{1}{m^2}$. By Claim 3.8, $\Pr[C_S^i] \geq 1 - e^{-p|S|}$.

Assume towards a contradiction that for each bidder $i$ and $v_{-i}$, $|\mathcal{S}_i^{v_{-i}}| < \frac{e^{\frac{m}{n^2}}}{10n^2}$. By the union bound:

$$\Pr[O \bigwedge_{i,S} C_S^i] = 1 - \Pr[\overline{O} \bigvee_{i,S} \overline{C_S^i}] > 1 - (e^{-\frac{m}{300}} + n \cdot \frac{e^{\frac{m}{n^2}}}{10n^2} \cdot e^{-p|S|}) \geq 1 - (e^{-\frac{m}{300}} + n \cdot \frac{e^{\frac{m}{n^2}}}{10n^2} \cdot e^{-\frac{m}{n^2}}) > 1 - \frac{1}{n}$$

Thus, there exists some instance for which all the events defined above occur. Let the output of the algorithm on this instance be $(A_1, \ldots, A_n)$. Let $T$ be the set of indices $i$ for which $|A_i| \leq \frac{pm}{n}$ and let $B$ be the set of indices for which $|A_i| > \frac{pm}{n}$. The welfare of the solution produced by the algorithm is:

$$\Sigma_{i \in T} v_i(A_i) + \Sigma_{i \in B} v_i(A_i) \leq n \cdot \frac{pm}{n} + \Sigma_{i \in B}(\frac{2|A_i|}{n} + \frac{1}{m^2}) \leq \frac{m}{n} + \frac{2m}{n} + \frac{n}{m^2} \leq \frac{4m}{n}$$

where for the first inequality we use the size of $A_i$'s in $T$ as an upper bound to their contribution, and use the fact that event $C_S^i$ occurs to bound the contribution of $A_i$'s in $B$. The second inequality holds since $(A_1, \ldots, A_n)$ is a an allocation, and thus $|\cup A_i| \leq m$. The leftmost inequality holds since $m \geq n^2$ (if $m < n^2$ the statement of the theorem guarantees that $A$ makes at one query, which is trivially true for every algorithm with a finite approximation ratio).

Since event $O$ occurs in this instance, the approximation ratio provided by the algorithm is at least $\frac{m(1 - \frac{1.01}{e})}{\frac{4m}{n}} > \frac{n}{10}$. A contradiction to the guaranteed approximation ratio. □

Now we are finally ready to define the structured submenu with the required size. Take $\mathcal{S}_{v_{-i}}$ of size $t = \frac{e^{\frac{m}{n^2}}}{10n^2}$, as guaranteed by the claim. Put the bundle $S \in \mathcal{S}_i^{v_{-i}}$ in bin $(k, x)$ if $|S| = k$ and $x \cdot m^{\frac{1}{5}} \leq p_S(v_{-i}) < (x+1) \cdot m^{\frac{1}{5}}$, where $x$ is an integer. There are $m^6$ bins, since for each $S \in \mathcal{S}_{v_{-i}}$ we have that $0 \leq p_S(v_{-i}) \leq m$ and each bundle is of size at most $m$. Let $\mathcal{S}$ be the set of size at least $\frac{t}{m^6}$ that consists of all bundles in the most congested bin. Notice that $\mathcal{S}$ is a structured submenu. This follows by Claim 3.6 and because all bundles in $\mathcal{S}$ are in the same bin: the price difference between every two bundles in the same bin is at most $m^{\frac{1}{5}}$ and all bundles in the same bin have the same size.

## 3.2 The Optimization Lemma

**Lemma 3.10 (Optimization)** *Let A be a truthful algorithm for combinatorial auctions with submodular bidders. Let $\mathcal{S} \subseteq \mathcal{R}_{v_{-i}}$ be a structured submenu, for some $v_{-i}$. Then, the number of value queries A makes is at least $|\mathcal{S}| - 1$.*

Denote the size of all sets in $\mathcal{S}$ by $k$. Let $t$ be greater than $2^m \cdot m$. For every $S^* \in \mathcal{S}$, define the following valuation $v_i^{S^*}$ of bidder $i$:

$$v_i^{S^*}(S) = \begin{cases} |S| \cdot t, & |S| < k; \\ k \cdot t - \frac{1}{m^4}, & S \in \mathcal{S} \text{ and } S \neq S^*; \\ k \cdot t, & S = S^* \text{ or } \exists T \in \mathcal{S} \text{ s.t. } S \text{ strictly contains } T; \\ t \cdot (k - \frac{1}{2^{m-|S|}}), & \text{otherwise.} \end{cases}$$

**Claim 3.11** *For every $S^*$, $v_i^{S^*}$ is non-decreasing and submodular.*

**Proof:** One can easily verify that $v_i^{S^*}$ is non-decreasing. We now show that all marginal values are non-increasing, hence the valuation is submodular. I.e., $v_i^{S^*}(S \cup \{j\}) - v_i^{S^*}(S) \leq v_i^{S^*}(T \cup \{j\}) - v_i^{S^*}(T)$, for every $T \subseteq S, j \notin T$. We divide the analysis into several simple cases:

- $|S \cup \{j\}| \leq k$: For every $T$, we have that $v_i^{S^*}(T \cup \{j\}) - v_i^{S^*}(T) = t$. On the other hand, $v_i^{S^*}(S \cup \{j\}) - v_i^{S^*}(S)$ equals to either $t$ (if $S \cup \{j\} = S^*$ or $|S \cup \{j\}| < k$), $t - \frac{1}{m^4}$ or $t(1 - \frac{1}{2^{m-|S|}})$ (in the second and fourth cases in the definition of $v_i^{S^*}$).

- $|S \cup \{j\}| > k$: by the previous bullet we are left with considering bundles $T$ such that $|T| \geq k-1$. If $v_i^{S^*}(T \cup \{j\}) - v_i^{S^*}(T) = k \cdot t$ then $v_i^{S^*}(S \cup \{j\}) - v_i^{S^*}(S) = 0$, which implies that the marginal value is non increasing. If $v_i^{S^*}(T \cup \{j\}) = k \cdot t - \frac{1}{m^4}$, then $v_i^{S^*}(T \cup \{j\}) - v_i^{S^*}(T) = t - \frac{1}{m^4}$. Since the maximum value of $v_i^{S^*}$ is $k \cdot t$, we have that the marginal value is non increasing in this case.

  The last case we have to consider is when $v_i^{S^*}(T \cup \{j\}) - v_i^{S^*}(T) = t \cdot (k - \frac{1}{2^{m-|T \cup \{j\}|}})$. Consider adding items from $(S \setminus T) \cup \{j\}$ one after the other in some arbitrary order. The marginal value of any additional item is either half of the value of the previous item (if the value of the new bundle is determined according to the fourth case in the definition of $v_i^{S^*}$), or exactly the marginal value of the previous item (if this is the first bundle for which the value is $k \cdot t$) or 0 (if there was a previous bundle with value $k \cdot t$). In either cases the marginal value decreases, as needed.

□

Below we show that when bidder $i$'s valuation is $v_i^{S^*}$ and the other bidders' valuations are $v_{-i}$, $S^*$ is his profit maximizing bundle. This implies that bidder $i$ must be allocated the bundle $S^*$. However, we show that finding $S^*$ cannot be done efficiently:

**Claim 3.12** *Finding $S^*$ requires $|\mathcal{S}| - 1$ value queries.*

**Proof:** Let $S'^*$ be such that $|S^*| = |S'^*|$. Observe that the valuations $v_i^{S^*}$ and $v_i^{S'^*}$ differ only in their value for $S^*$ and $S'^*$. Thus, a query for the value of a bundle $S$ only tells us whether the valuation is $v_i^S$ or not. In the worst case, we have to query the value of every bundle $S \in \mathcal{S}$ (except the "last" bundle) to determine $S^*$. □

We are left with showing that when bidder $i$'s valuation is $v_i^{S^*}$ then $S^*$ is his profit-maximizing bundle (notice that $v_i^{S^*} - p_{v_{-i}} > 0$). The proof consists of the following series of simple claims.

**Claim 3.13** $v_i^{S^*}(S^*) - p_{v_{-i}}(S^*) > v_i^{S^*}(S) - p_{v_{-i}}(S)$, for every bundle $S$ such that $v_i^{S^*}(S) \leq t \cdot (k - \frac{1}{2^m})$.

**Proof:** It suffices to show that
$$k \cdot t - p_{v_{-i}}(S^*) > t \cdot (k - \frac{1}{2^m}) - 0$$
which holds if $\frac{t}{2^m} > p_{v_{-i}}(S^*)$. The claim now follows since by the properties of a structured submenu $p_{v_{-i}}(S^*) \leq m$. □

**Claim 3.14** $v_i^{S^*}(S^*) - p_{v_{-i}}(S^*) > v_i^{S^*}(S) - p_{v_{-i}}(S)$, for every bundle $S$ such that $|S| > k$ where there exists some $T \in \mathcal{S}$ such that $T \subseteq S$.

**Proof:** Observe that $v_i^{S^*}(S) = v_i^{S^*}(S^*) = t \cdot k$. To finish the proof we show that $p_{v_{-i}}(S) > p_{v_{-i}}(S^*)$. By the properties of structured submenu, since $S$ contains some set in $\mathcal{S}$, we have that $p_{v_{-i}}(S) > p_{v_{-i}}(T) + \frac{1}{m^3}$. We also have that $|p_{v_{-i}}(S) - p_{v_{-i}}(S^*)| < \frac{1}{m^5}$. This implies that $p_{v_{-i}}(S) - p_{v_{-i}}(S^*) > \frac{1}{m^3} - \frac{1}{m^5}$, as needed. □

**Claim 3.15** $v_i^{S^*}(S^*) - p_{v_{-i}}(S^*) > v_i^{S^*}(S) - p_{v_{-i}}(S)$, for every bundle $S \neq S^*$ such that $|S| = k$.

**Proof:** $v_i^{S^*}(S^*) = t \cdot k$. By the previous claims we are left with the case where $v_i^{S^*}(S) = t \cdot k - \frac{1}{m^4}$. By the properties of structured submenus $|p_{v_{-i}}(T) - p_{v_{-i}}(S^*)| < \frac{1}{m^5}$. The claim follows. □

## 4 Flexible Combinatorial Public Projects

We show that every randomized universally truthful algorithm for flexible combinatorial public projects that achieves an approximation ratio of $m^{\frac{1}{2}-\epsilon}$ requires exponential number of value queries. The proof is a simpler version of the result for combinatorial auctions with submodular bidders. We highly recommend reading Section 3 first. We prove the result for deterministic mechanisms. A lower bound of $m^{\frac{1}{2}-\epsilon}$ for randomized universally truthful mechanisms may be obtained as in Section A.1.

**Theorem 4.1** Let $A$ be a deterministic truthful $m^{\frac{1}{2}-\epsilon}$-approximation mechanism for flexible combinatorial public projects, for some constant $\epsilon > 0$. Then, $A$ makes at least $\frac{e^{m^{2\epsilon}}}{100m^6}$ value queries.

From now on let the number of items selected in the problem to be $\sqrt{m}$. The proof consists of the following lemmas.

**Lemma 4.2** Let $A$ be a truthful algorithm for extended combinatorial public projects with an approximation ratio of $\frac{1}{m^{\frac{1}{2}-\epsilon}}$, for some constant $\epsilon > 0$. Then, there exists $v_{-i}$, $\mathcal{S}$, $|\mathcal{S}| \geq \frac{e^{-m^{2\epsilon}}}{100m^6}$, such that $\mathcal{S} \subseteq \mathcal{R}_{v_{-i}}$ is a structured submenu.

**Proof:** We prove the result for the special case where we fix some bidder $i$ and set all $v \in v_{-i}$ to be identically zero (i.e., for every $S$ and $v \in v_{-i}$, $v(S) = 0$). A random polar-additive valuation $v$ is now constructed as follows: for each item $j$, set independently at random $v_i(j) = 1$ with probability $p = \frac{1}{\sqrt{m}}$ or $v_i(j) = \frac{1}{m^3}$ with probability $1 - p$.

**Definition 4.3** *Fix bidder $i$. Let $v_{-i}$ be the set of valuations that are all identically zero. Let $\mathcal{S} = \{S | \exists$ a polar-additive valuation $v_i$ s.t. $A(v_i, v_{-i})_i = S\}$.*

**Claim 4.4** *For every bidder $i$ and $v_{-i}$, such that each $v \in v_{-i}$ is identically $0$, $|\mathcal{S}| > \frac{e^{m^{2\epsilon}}}{100}$.*

**Proof:** Let $C_S$ be the event where $v(S) < m^\epsilon$ for each bundle $S \in \mathcal{S}$ where $v$ is a random polar-additive valuation and $\epsilon > 0$ is constant. By the chernoff bounds and since $v$ is polar additive, $\Pr[v(S) \geq m^\epsilon] \geq 1 - e^{-m^{2\epsilon}}$, where we use the fact that the probability is minimized when $|S| = \sqrt{m}$, the maximum possible bundle size that can be selected in the problem.

Let $O$ be the event where $\max_{S:|S|=\sqrt{m}} v(S) \geq \frac{\sqrt{m}}{2}$. By the chernoff bounds, $O$ occurs with probability at least $1 - e^{-m^{\frac{1}{4}}}$. Assume towards a contradiction that $|\mathcal{S}| \leq \frac{e^{m^{2\epsilon}}}{100}$. By the union bound:

$$\Pr[O \bigwedge_S C_S] = 1 - \Pr[\overline{O} \bigvee_S \overline{C_S}] > 1 - (e^{-m^{\frac{1}{4}}} + \frac{e^{m^{2\epsilon}}}{100} \cdot e^{-m^{2\epsilon}}) > \frac{49}{50}$$

Thus there is an instance where all the events defined above occur simultaneously. Let $S$ be the set that the algorithm outputs in this instance. By definition $S \in \mathcal{S}$. Since the event $C_S$ occurs we have that the welfare that the algorithm provides is at most $m^\epsilon$. On the other hand, the optimal solution has a value of at least $\frac{\sqrt{m}}{2}$, since the event $O$ occurs. Thus the algorithm provides an approximation ratio worse than $m^{\frac{1}{2}-\epsilon}$ for this instance, a contradiction. □

We now specify the structured submenu with the required size. Take $\mathcal{S}$ of size $t = \frac{e^{m^{2\epsilon}}}{100}$, as guaranteed by the claim. Put the bundle $S \in \mathcal{S}$ in bin $(k, x)$ if $|S| = k$ and $x \cdot m^{\frac{1}{5}} \leq p_S(v_{-i}) \leq (x+1) \cdot m^{\frac{1}{5}}$, where $x$ is an integer. There are $m^6$ bins, since for each such $S$ we have that $0 \leq p_S(v_{-i}) \leq m$ and the size of each bundle is at most $m$. Let $T$ be the set of size at least $\frac{t}{m^6}$ that consists of all bundles in the most congested bin. Notice that $T$ is a structured submenu. This follows by Claim 3.6 and because all bundles in $T$ are in the same bin: the price difference between each two bundles in $T$ is $m^{\frac{1}{5}}$ and all bundles in $T$ have the same size. □

The proof of the following lemma is identical to the proof of Lemma 3.10:

**Lemma 4.5 (Optimization)** *Let $A$ be a truthful algorithm for extended combinatorial public projects. Let $\mathcal{S} \subseteq \mathcal{R}_{v_{-i}}$ be a structured submenu, for some $v_{-i}$. $A$ makes at least $|\mathcal{S}| - 1$ value queries.*

## 5 Truthful in Expectation Mechanisms for Exact CPP

This section shows that any truthful in expectation mechanisms for exact combinatorial public projects that guarantees an approximation ratio better than $m^{\frac{1}{2}-\epsilon}$ requires exponential number of value queries. This is the first lower bound on the power of polynomial time truthful in expectation mechanisms in any setting. We fix the number of items selected in the problem to be $\sqrt{m}$. We observe that in a truthful in expectation mechanism, prices are given to distributions and not to bundles[7].

**Theorem 5.1** *Let $A$ be a truthful in expectation mechanism for exact combinatorial public projects with an approximation ratio of $m^{\frac{1}{2}-\epsilon}$, for some constant $\epsilon > 0$. $A$ makes at least $e^{m^{2\epsilon}}$ value queries.*

The proof of this theorem can be found in the appendix.

---
[7]Note that the natural way of normalizing prices, by setting $p_{v_{-i}}(\emptyset) = 0$ does not make sense in combinatorial public projects, where the empty bundle is never selected. Thus we set $p_{v_{-i}}(D) = 0$, where $D$ is the distribution that the algorithm outputs when the valuation of bidder $i$ is identically zero.

**Acknowledgments**

I would like to thank Itai Ashlagi, Hu Fu, Bobby Kleinberg, Noam Nisan, Sigal Oren, and Michael Schapira for helpful discussions and comments.

## A  Missing Sections and Proofs

### A.1  Impossibility Results for Randomized Universally Truthful Mechanisms

We briefly sketch how to obtain impossibility results for randomized universally truthful randomized mechanisms. We use ideas introduced in [8, 7].

**Theorem A.1** *Let $A$ be a randomized universally truthful $\frac{n}{10}$-approximation mechanism for combinatorial auctions with submodular valuations. Then, $A$ makes at least $\frac{e^{\frac{m}{n^2}}}{10n^2 \cdot m^8}$ value queries in expectation.*

Instead of working with randomized mechanisms that provide a good approximation ratio on all input, it will be easier to work with deterministic mechanisms that provide a good approximation ratio on "many" inputs. A reduction in this spirit can be obtained as follows:

**Definition A.2** *Fix $\alpha \geq 1$, $\beta \in [0,1]$, and a finite set $U$ of instances of combinatorial auctions with submodular bidders. A deterministic algorithm $B$ is $(\alpha, \beta)$-good on $U$ if $B$ returns an $\alpha$-approximate solution for at least a $\beta$-fraction of the instances in $U$.*

**Proposition A.3 (essentially [8, 7])** *Let $U$ be some finite set of instances, and let $\gamma > \alpha \geq 1$. Let $A$ be a universally truthful mechanism that provides an expected welfare of $\frac{OPT(I)}{\alpha}$ for every instance $I \in U$ with expected number of value queries $val(A)$. Let $\alpha' = \frac{1}{\frac{1}{\alpha} - \frac{1}{\gamma}}$. Then, for every $\tau > 1$ there is a (deterministic) algorithm in the support of $A$ that is $(\alpha'\tau, (1 - \frac{1}{\tau})/\alpha')$-good on $U$ and and makes $\gamma \cdot val(A)$ value queries.*

We would like to prove that there is every universally truthful randomized $\frac{n}{10}$-approximation mechanism for combinatorial auctions with submodular valuations must make exponential number of value queries. From the proposition, using $\gamma = m, \tau = 2$ and $U$ is the set of all instances where every valuation is polar additive, we have that there exists a $(\frac{\alpha m}{m-\alpha}, \frac{m-\alpha}{2\alpha m})$-good algorithm $A'$ on $U$ that makes $m \cdot val(A)$ value queries. We will a lower bound on the number of value queries $A'$ must make, hence we also bound the number of queries $A$ makes. This will conclude the proof.

We show the existence of an exponentially large structured submenu in $A'$. We modify the proof of Lemma 3.3 as follows. Let $W$ be the event that the $A'$ provides an approximation ratio of $\frac{\alpha m}{m-\alpha}$ on the random instance. Observe that since $A'$ is $(\frac{\alpha m}{m-\alpha}, \frac{m-\alpha}{2\alpha m})$-good, $W$ occurs with probability at least $\frac{m-\alpha}{2\alpha m}$. In particular, for $\alpha = \frac{n}{10}$, $\Pr[W \wedge O \wedge_{i,S} C_S^i] > 0$. The proof now continues as before to show that $A'$ has a structured submenu of size $\frac{e^{\frac{m}{n^2}}}{10n^2 \cdot m^6}$. Lemma 3.10 shows that $\frac{e^{\frac{m}{n^2}}}{10n^2 \cdot m^6}$ is a lower bound on the number of value queries that $A'$ makes. This implies that the number of value queries of the randomized algorithm $A$ is as specified.

## A.2 Proof of Theorem 5.1

The proof consists of the following lemmas.

**Lemma A.4** *Let $A$ be a truthful algorithm with an approximation ratio of $m^{\frac{1}{2}-\epsilon}$, for some constant $\epsilon > 0$. Fix all bidders $v_{-i}$ but bidder $i$ to be identically zero. Let $S$ be some bundle of size $\sqrt{m}$. There exists a distribution $D$ in the menu of $i$ such that $\Pr_{S \sim D}[|S \cap T| > m^\epsilon] > \frac{1}{m^{\frac{1}{2}-\epsilon}}$ and $p_{v_{-i}}(D) \leq \sqrt{m}$.*

**Proof:** Let $v_i$ be the valuation of bidder $i$ with $v_i(\{j\}) = 1$ for every $j \in T$ and $v_i(\{j\}) = 0$ otherwise, and for every bundle $U$, $v_i(U) = \Sigma_{i \in U} v_i(\{j\})$. By the guaranteed approximation ratio, $A$ outputs a distribution $D$ with $\Pr_{T \sim D}[S \cap T \neq \emptyset] > \frac{1}{m^{\frac{1}{2}-\epsilon}}$. By individual rationality we must also have that $p_{v_{-i}}(D) \leq \sqrt{m}$. □

For a bundle $T$, $|T| = \sqrt{m}$ define the following submodular valuation, where $t = m^{10}$:

$$v_T(S) = \begin{cases} t \cdot |S|, & |S| < \sqrt{m}; \\ t \cdot \sqrt{m}, & |S \cap T| > m^\epsilon, |S| = \sqrt{m}; \\ t \cdot \sqrt{m}, & |S| > \sqrt{m}; \\ t \cdot (|S| + \frac{1}{2}), & \text{otherwise.} \end{cases}$$

Also define the following valuation $v_\emptyset$:

$$v_\emptyset(S) = \begin{cases} t \cdot |S|, & |S| < \sqrt{m}; \\ t \cdot \sqrt{m}, & |S| > \sqrt{m}; \\ t \cdot (|S| + \frac{1}{2}), & \text{otherwise.} \end{cases}$$

**Lemma A.5** *Every algorithm that uses value queries only makes $e^{m^{2\epsilon}}$ value queries in expectation to distinguish $v_T$ from $v_\emptyset$.*

**Proof:** Choose $T$ uniformly at random among all bundles of size $\sqrt{m}$. Observe that for all bundles $S$, $|S| \neq \sqrt{m}$, $v_\emptyset(S) = v_T(S)$. Thus from now on we only count queries to bundles of size $\sqrt{m}$. Fix some bundle $S$, $|S| = \sqrt{m}$. It holds that $\Pr[v(S) \neq (\sqrt{m} + \frac{1}{2}) \cdot t] = \Pr[S \cap T > m^\epsilon] < e^{-m^{2\epsilon}}$, where the probability is taken over the random choice of $T$. Thus, for every deterministic mechanism there is some $T$ for which $A$ makes at least $e^{m^{2\epsilon}}$ queries. Furthermore, by Yao's principle every randomized mechanism makes that number of queries in expectation. $\square$

And finally:

**Lemma A.6** *Let $A$ be a truthful in expectation mechanism that provides an approximation ratio of $m^{\frac{1}{2}-\epsilon}$ and makes at most $val(A)$ value queries. Then, there is a mechanism that makes $poly(val(A))$ value queries that distinguishes $v_T$ from $v_\emptyset$, for every $T$ with high probability.*

**Proof:** Let $D$ be the distribution that the algorithm outputs when the valuation of bidder $i$ is $v_T$. Fix some bundle $T$, $|T| = \sqrt{m}$. Let $D_T$ be the distribution in the menu guaranteed by Lemma A.4. We now show that $\Pr_{S \sim D}[|S \cap T| > m^\epsilon] > \frac{1}{m^4}$. Suppose not. Notice that

$$E_{S \sim D_T}[v_T(S) - p_{v_{-i}}(D_T)] \geq \Pr_{S \sim D_T}[|S \cap T| > m^\epsilon](\sqrt{m} \cdot t) + \Pr_{S \sim D_T}[|S \cap T| \leq m^\epsilon]((\sqrt{m} - \frac{1}{2}) \cdot t) - m$$

$$\geq \frac{1}{m^{\frac{1}{2}-\epsilon}}(\sqrt{m} \cdot t) + (1 - \frac{1}{m^{\frac{1}{2}-\epsilon}})((\sqrt{m} - \frac{1}{2}) \cdot t) - \sqrt{m}$$

Where the first inequality holds since every bundle in the support of $D_T$ is of size $\sqrt{m}$, according to the definition of exact combinatorial public projects. However, a similar calculation yields that

$$E_{S \sim D}[v_T(S) - p_{v_{-i}}(D)] \leq \Pr_{S \sim D}[|S \cap T| > m^\epsilon](\sqrt{m} \cdot t) + \Pr_{S \sim D}[|S \cap T| \leq m^\epsilon]((\sqrt{m} - \frac{1}{2}) \cdot t) - 0$$

$$\leq \frac{1}{m^4}(\sqrt{m} \cdot t) + (1 - \frac{1}{m^4})((\sqrt{m} - \frac{1}{2}) \cdot t) - 0$$

I.e., $D$ is less profitable than $D_T$, a contradiction since the algorithm outputs the distribution $D$. Thus, when bidder $i$'s valuation is $v_T$, $A$ outputs some distribution $D$ with $\Pr_{S \sim D}[|S \cap T| > m^\epsilon] \geq \frac{1}{m^4}$.

If we run $A$ polynomially many times (i.e., sample from $D$ polynomially many times), we will be able to find some set $S$, $|S| = \sqrt{m}$, such that $|S \cap T| > m^\epsilon$ with high probability. Hence we can distinguish $v_T$ from $v_\emptyset$, as needed. $\square$